\documentclass[fleqn,11pt,a4paper]{article}

\usepackage{epsfig,amsmath,amssymb}

\begin{document}


\def\REF#1{(\ref{#1})}

\def\t{\tau}
\def\dt#1{{\dot{#1}}}
\def\td#1{{\widetilde{#1}}}
\def\ww{{{\bf W}}}
\def\KK{{\cal K}}
\def\UU{{\cal U}}
\def\II{{\cal I}}
\def\OO{{\cal O}}
\def\FF{{\cal F}}
\def\EE{{\cal E}}
\def\ACC{{\cal A}}
\def\gg{{\cal G}}
\def\CCC{{\cal C}}

\def\CC#1#2{{\cal C}^#1{}_#2\,}
\def\AA#1#2{{\cal A}^#1{}_#2\,}
\def\LL#1#2{{\Lambda}^#1{}_#2\,}
\def\ppp#1{{p}^#1{}\,}
\def\PHH#1{{\cal Q}^#1\,}
\def\YY{{\cal Y}}

\def\hCC#1#2{\hat{\cal C}^#1{}_#2\,}
\def\hAA#1#2{\hat{\cal A}^#1{}_#2\,}
\def\hPHH#1{\hat{\cal Q}^#1\,}
\def\hYY{{\hat{\cal Y}}}

\def\ff#1{{{\cal F}}^#1\,}
\def\hff#1{\hat{{\cal F}}^#1\,}

\def\SU{{\cal M}}
\def\SSS{{\cal S}}
\def\wdg{\wedge}
\def\REF#1{(\ref{#1})}

\def\CITE#1{\cite{#1}}

\def\DOT#1#2{{#1 \cdot #2}}

\def\CROSS#1#2{{#1 \times #2}}

\def\bfA{{\bf A}}

\def\ee{{\cal E}}
\def\sstar{{\bf \#}}
\def\tt{{(\tau)}}

\def\hRR#1#2#3#4{\hat{R}^{#1}{}_{#2#3#4}\,}
\def\RR#1#2#3#4{R^{#1}{}_{#2#3#4}\,}
\def\hTT#1#2#3{\hat{T}^{#1}{}_{#2#3}\,}
\def\TT#1#2#3{T^{#1}{}_{#2#3}\,}
\def\hFF#1#2{{\hat F^{#1#2}}}

\def\pprn#1{\,\partial^{#1}_\lambda\,}
\def\PP{{\prime\prime}}
\def\PPP{{\prime\prime\prime}}
\def\se#1{{\star\, e^{#1}}}
\def\bfr{{\bf r}}

\def\pp{\partial}
\def\L{\Lambda}
\def\l{\lambda}
\def\a{\alpha}
\def\b{\beta}
\def\d{\delta}
\def\g{\gamma}
\def\s{\sigma}
\def\nn{{\cal W}}
\def\SS{\Sigma}
\def\S{\UU}

\def\bfs{{\bf s}}

\def\bfh{{\bf h}}

\def\bfE{\hat{\bf E}}

\def\bfB{\hat{\bf B}}

\def\bfA{{\bf A}}

\def\bfP{{\bf P}}

\def\dd{\, d \,}

\def\modsq#1{{\vert #1 \vert^2}}

\def\mat{\Phi}
\def\LX{{\cal L}_X}
\def\gg#1{\delta_{#1}}
\def\hX#1{\hat{X}_#1}

\def\be#1{{\begin{equation}#1\end{equation}}}
\def\aa{A}
\def\bb{B}
\def\aaa{{A^\prime}}
\def\bbb{{B^\prime}}

\title{\bfseries Field-Particle Dynamics in Spacetime Geometries}
\author{\bfseries Robin W. Tucker\thanks{Department of Physics, Lancaster
University, UK (email : r.tucker@lancaster.ac.uk)} } \maketitle


\begin{abstract}

 With the aid of a Fermi-Walker chart associated with an orthonormal
 frame attached to a time-like curve in spacetime, a discussion
 is given of relativistic balance laws that may be used to
 construct models of massive particles with spin, electric charge
 and a  magnetic moment, interacting with background electromagnetic
  fields and gravitation described by non-Riemannian geometries.
  A natural generalisation to relativistic Cosserat media is immediate.
\end{abstract}

\section{Introduction}

The language of relativistic fields on a spacetime offers,
probably,
 the most succinct and complete description of the basic laws of
physics. From these laws we expect to accommodate the enormous
amount of data derived from experiment, predict new phenomena and
modify or reject laws in accordance with observation. Of necessity
this process invariably involves some level of approximation. As
experimental technology improves it becomes possible to refine
such approximations and thereby seek more detailed experimental
verification of theoretical predictions.

A field description of relativistic gravitation was formulated by
Einstein almost a century ago.  Since then it has proved one of
the most successful descriptions of large scale gravitational
phenomena in all of physics  and has profound implications for the
notions of space and time on all scales. Nevertheless this theory
of gravitation cannot be complete and it poses many challenges in
both classical and quantum domains. Modern experimental tools of
increasing sophistication are now being brought to bear on some of
these problems with the search for classical gravitational waves,
the breakdown of the \lq\lq equivalence principle" and the
influence of gravitation on the \lq\lq collapse of the
wave-function". Along with these technological advances new
theoretical avenues continue to be explored in which all the basic
interactions between matter are subsumed into a coherent
framework.

Viable approximation schemes that link the results of experimental
probes and  theoretical descriptions of the basic interactions
will remain as  bridges between these continuing developments.
Since the early days of relativity it was recognised that the
derivation of the motion of extended \lq\lq matter" in
gravitational fields was a non-trivial problem and great effort
was expended in deriving effective approximations. The work of
Einstein, Infeld, Hoffman, Mathisson, Papapetrou, Pryce ,
Tulczyjew and others \CITE{mat,Papapetrou:1951, pryce, Tul,weyss}
culminated in a series of papers by Dixon, \CITE{dixon1:1964,
dixon1:1965,dixon1:1970, dixon2:1970, dixon3:1973, dixon4:1974},
Ehlers and Rudolph \CITE{ehlers:1977} and Kunzle \CITE{kunz},
where it was demonstrated that a viable scheme could be
established, given suitable subsidiary conditions, in which an
exact field description could be replaced by an approximation
procedure in which extended matter was regarded at each order of
approximation by a finite set of (non-unique) matter and
electromagnetic multipoles.  Such a scheme offers in principle a
viable approach to the motion of extended matter and deserves to
be taken seriously in the context of modern experiments that
purport to analyse the fundamentals of the gravitational
interaction with spinning matter. However theories of gravitation
with spinning matter sit somewhat un-naturally in the
pseudo-Riemannian (i.e. torsion free) environment in which the
whole Dixon framework was formulated. Furthermore it is not
necessary to have \lq\lq spinning sources" to accommodate
gravitational fields with torsion. Indeed a Lorentz
gauge-covariant formulation of the Brans-Dicke theory naturally
yields a metric-compatible connection on the bundle of linear
frames over spacetime with torsion determined by the gradient of
the Brans-Dicke scalar field \CITE{dereli_tucker_tor}. Since many
modern variants of this theory (and various \lq\lq string
theories") abound with scalar gravitational fields one should
seriously consider the possibility that gravitation may have a
torsional component that is absent (by hypothesis) in the
pseudo-Riemannian framework given by Einstein.

The multipole analysis formulated by Dixon relies heavily on the
use of bi-tensors and tensor densities and the pseudo-Riemannian
structure of spacetime. To extend this methodology to general
spacetime structures a new approach is adopted here based on the
use of differential forms, structure equations, a (generalised)
Fermi chart associated with a timelike worldline and the transport
of a Fermi-Walker frame along spacelike autoparallels of a
metric-compatible connection with torsion. This approach may be
contrasted with  the use of adapted coordinates employed by
Tulczyjew  and Tulczyjew  \CITE{polish_book} in a torsion-free
context. Although they employ a Fermi chart in discussing the
equations of motion of an electrically neutral particle, their
methods, detailed calculations and conclusions differ
significantly from the results to be discussed below even in the
case considered, where the torsion is zero. In particular their
resulting equations admit unnatural solutions in Minkowski
spacetime as will be discussed in section VIII.

The  approach to be considered here considers the modelling of
relativistic massive particle interactions with background
electromagnetic and gravitational fields with possible torsion. As
stressed above it is assumed that the {\it underlying} laws
describing all phenomena can be formulated in terms of tensor and
spinor {\it fields} over a manifold. Depending on the physical
scales involved such fields are in general subject to either the
laws of classical or quantum field theory. In this paper emphasis
is  on classical fields. In particular it is assumed that
classical gravitation can be described in terms of a geometry on a
spacetime and the electromagnetic field is  a closed 2-form on
this manifold. Like Newtonian gravitation, electromagnetism  is
distinguished from other gauge fields by its long range character.
It can also be related to the curvature of a connection on a
principle $U(1)$ bundle over spacetime. The connection on the
bundle of linear frames over spacetime will be that induced from
the bundle or orthonormal frames. Thus the local gauge group for
gravitation may be regarded as the covering of the Lorentz group
$SO(1,3)$. Shorter range 2-form (internal symmetry non-Abelian
Yang-Mills) fields, regarded as curvatures associated with
connections on other principle group bundles are also responsible
for the basic interactions in Nature. Matter fields (regarded as
sections of associated bundles) carry representations of these
Yang-Mills groups and their gauge covariant interactions with
gauge fields are responsible for the \lq\lq charges" of massive
field quanta. All fields influence the geometry of spacetime and
in turn the gravitational field influences the behavior of all
other fields. The field laws for the coupled system of gauge
fields and matter may be compactly derived as differential
equations that render extremal the integral of some (action)
4-form on spacetime. This integral is constructed to be invariant
under the action of all local gauge groups. The resulting
classical variational field equations are thereby covariant under
changes of section induced by associated local changes of frame
(gauge). In this article the Yang-Mills internal symmetry group
made explicit will be restricted to electromagnetism. It is the
gauge invariance of some action integral that is of paramount
importance in the developments below.

\section{Balance Laws}

Classical {\it point particles} are an abstraction. In general
they offer a useful approximation to the behavior of localised
matter and form the basis of rigid body dynamics in Newtonian
physics following the laws of Newtonian particle dynamics. The
behavior of Newtonian deformable continua require new laws that
exploit the high degree of Killing symmetry in spacetime devoid of
gravitation where matter, (observed from weakly accelerating
reference frames) has a small speed relative to that of light. The
{\it balance laws} of energy, linear and angular momentum  must be
separately postulated in order to provide a framework for the
description of non-relativistic deformable continua. However the
evolution of such continua can only be determined from such
balance laws once a choice of constitutive relations (Cauchy
stress tensor) consistent with the laws of thermodynamics has been
made. Such relations serve to render the balance laws predictive
and are either phenomenological in nature or arise as
coarse-grained approximations applied to a statistical or quantum
analysis that recognises the atomistic nature of matter. Newtonian
balance laws refer to arbitrary distributions of matter in space.
Point particle and rigid-body dynamics arise as zeroth and first
order multipole approximations \CITE{dixon_book}.  There are other
types of approximation that are useful when the continuum is
slender in either one or two dimensions relative to the third
(i.e. rods and shells). Furthermore the internal state of such
structures can be
 expressed as degrees of freedom associated with a field of
frames in the medium (Cosserat models). In all such cases a
successful phenomenological description is sought in terms of a
small number of physical parameters (mass, moment of inertia
tensor, elastic moduli, pressure, density, thermal and viscosity
coefficients...) that is amenable to experimental verification.
For extended structures, such as fluids or localised structures
such as strings or rods it is the relatively weak interaction of
the continuum with its environment that renders such a description
effective and enables one to distinguish externally prescribed
forces and constraints from the responses that they produce.

An analogous strategy is sought to describe the {\it relativistic}
motion of localised matter from an underlying field theory
involving gravitation and electromagnetism. Since the matter
fields may carry representations of the Lorentz group (or its
covering) as well as $U(1)$ one expects  to accommodate
electrically charged particles with some kind of spin. The basic
approximation considered here  treats the gauge fields and
gravitation as a background that is prescribed independently of
the matter fields. These on the other hand are required to satisfy
gauge-covariant field equations in the presence of the prescribed
gauge fields and gravitation. The metric of the background
spacetime  will, however, possess no particular isometries to
provide the analogues of the Newtonian balance laws. Nevertheless
the principle of local gauge covariance for the interactions
between fields provides powerful dynamical constraints on the
field variables that arise in any variational description. This
observation is well known and its implications can be found in
many parts of the literature. Such constraints will be explored
below in the context of  adapted orthonormal frame fields and
associated coordinates based on a  Fermi chart associated with a
time-like curve in spacetime.

 The  equations of motion proposed by Dixon offer a
consistent dynamical scheme \cite{ehlers:1977} for the classical
behavior of \lq\lq \lq\lq spinning matter"
\cite{wald:1972,wald:1974} in pseudo-Riemannian spacetimes. In the
lowest pole-dipole approximation  they can be solved analytically
for electrically neutral  particles in gravitational fields with
high symmetry \cite{tod:1976} but in general recourse to numerical
methods is required \cite{rwt:2001}, \CITE{Semerak, driven}. We
shall recover these equations below in spaces with zero torsion
albeit without recourse to bi-tensors and tensor densities.
Furthermore we shall show that in background spacetimes in which
gravitation is described by geometries having connections with
torsion, electrically charged spinning particles  deviate from
such histories due to prescribed interactions between the
background fields and the
 spin and charge of the particle.

We argue that in more general spacetimes even electrically neutral
spinless particles may have histories that follow autoparallels of
connections with torsion instead of geodesics of the spacetime
metric.

\section{Fermi Coordinates}

Let $M$ be a smooth manifold with a metric tensor $g$ and a metric
compatible connection $\nabla$. Denote the tangent space to $M$ at
point $p$ by $T_pM$. For any $p$ and any vector $v$ in $T_pM$ let
$\g_v:\l\mapsto \g_v(\lambda)\in M$ be the unique autoparallel
($\nabla _{\g_v{}^\prime}{\g_v{}^\prime }=0$)   of $\nabla$ with
\be {\g_v(0)=p} \be{ \g^\prime_v(0)=v.} Provided $\g_v(1)$ exists
denote it by $\exp_p(v)$ i.e. \be {\exp_p(v)\equiv \g_v(1) \in M.}
The exponential map $\exp_p: T_pM \mapsto M$ is therefore defined
in a neighbourhood of $0\in T_pM$ and is in fact a diffeomorphism
\CITE{sakai,RWT_book} .

Suppose $M$ is a four dimensional spacetime and  $g$ has
Lorentzian signature $(-,+,+,+)$. The geometry of $M$ is described
by $g$ and $\nabla$ which is metric-compatible but not assumed to
be torsion free. Let $\s:t\in {\bf R^+} \mapsto M$ be a time-like
future pointing affinely parametrised curve
($g(\dot\s,\dot\s)=-1$) and $\hat\FF\equiv\{\hX0,\hX1,\hX2,\hX3\}$
a $g$-orthonormal frame on $\s$ with
$\dot\s(t)=\hX0\vert_{\s(t)}.$ Then  a Fermi chart $\Psi$ with
Fermi coordinates $(t,x^1,x^2,x^3)$  relative to the curve $\s$,
frame $\hat\FF$ and arbitrary point $\s(t_0)$ on the image of $\s$
is defined by: \be{\Psi^0(\exp_{\s(t)} v)=t-t_0}
\be{\Psi^k(\exp_{\s(t)} v) =x^k} where $v=\sum_{j=1}^3 x^j\hX j
\in T_{\s(t)}M$. In the following repeated Latin indices will be
summed from 1 to 3 while repeated Greek indices will be summed
from 0 to 3. The chart is defined on any open neighbourhood $\S$
of $\s$ containing points that can be joined to $\s$ by an
affinely parameterised unique autoparallel meeting $\s$
orthogonally. Such a definition gives rise to a class of Fermi
charts for a given curve $\s$ with frames  related to each other
by local $SO(3)$ transformations and translations of origin
$\s(t_0)$ along $\s$ \CITE{tubes}. In any particular chart with
coordinates $\{t,x^k\}$ one has  the local vector fields
$\{\pp_t,\pp_k\equiv\frac{\pp}{\pp x^k}\}$ on $\S$ which by
construction are g-orthonormal on $\s$. Furthermore, by
definition, the space-like autoparallels $\g_{{}_{\pp_k}}$ that
leave $\s$ at $\{t,0,0,0\}$ can be affinely parameterised in this
chart by the equations \be{t-t_0=\t}\be{x^k=p^k\l} where
$(p^1){}^2+(p^2){}^2+(p^3){}^3=1$. Henceforth the origin is fixed
 with $t_0=0$. The parameters $p^k$ are the direction cosines of
the tangent to such an autoparallel, with respect to $\hat\FF$ at
$\{t,0,0,0\}$. It will prove useful below to introduce
Fermi-normal hyper-cylindrical polar coordinates $\{\t,\l,p^k\}$
where $(p^k\,p^k)=1$ on part of $\S$. These are naturally related
to the coordinates $\{t,x^k\}$ by the transformation $t=\t$ and
$x^k=p^k\l$. Thus any function $f$ of $t,x^1,x^2,x^3$ can be
regarded as a function  $F$ of $t,\l,p^1,p^2,p^3$ on a suitable
domain \be{f(t,x^1,x^2,x^3)=f(t,\l p^1,\l p^2, \l p^3) =
F(\t,\l,p^1,p^2,p^3).} Using the inverse relations
\be{\l^2(x^1,x^2,x^3)=x^j x^j} \be{
p^k(x^1,x^2,x^3)=\frac{x^k}{\l(x^1,x^2,x^3)}    } where $\l\ne 0$
one has the directional derivative \be{\partial_\l=p^k\pp_k} and
the relations $\pp_\l p^k=0$ and $\frac{\pp p^k}{\pp
p^j}=-\frac{p^j}{p^k}$ for $j\ne k$.
 If $F$ is
analytic in $\l$: \be{F(\t,\l,p^1,p^2,p^3)=\sum_{k=0}^\infty
\frac{\l^k}{k!}\,(\pp^k{}_\l F)(\t,0,p^1,p^2,p^3)}
 \be{{} =
\sum_{k=0}^\infty\,\,\sum_{i_1,i_2\ldots i_k=1}^3
\frac{\l^k}{k!}\, p^{i_1}\ldots p^{i_k} \,(\pp_{i_1}\ldots
\pp_{i_k} f) (t,0,0,0)} Such expansions will be used to represent
the structure 0-forms and 1-forms to be introduced below and
partial derivatives of $F$ with respect to $\l$ will be denoted
$F^{\prime}$. To eliminate excessive notation, expressions
evaluated on the curve $\s$ will be understood to be represented
in the coordinate system $\{t,x^k\}$ and the abbreviation $\hat f$
often used for $f$ restricted to $\s(t)$.

\section{Ortho-normal frames on $\S$}

The coordinate frame $\FF\equiv \{\pp_t,\pp_k\}$ on $\S$ is only
orthonormal on $\s$ where it coincides with $\hat\FF$. However one
can displace the frame $\hat\FF$ by parallel transport along the
space-like autoparallel curves that emanate from $\s$ using the
connection $\nabla$. Since this connection is assumed
metric-compatible this process will give rise to a field of
orthonormal frames $\OO$  on $\S$ \CITE{hicks}. The next task is
to calculate the orthonormal coframe field $\{e^\a\} $ dual to
$\OO$ in terms of the coframe $\{ d\,t,  d\, x^k\}$ dual to $\FF$.
If one assumes that the curvature and torsion are analytic in the
vicinity $\S$ of $\s$ this can be done by recursively
differentiating the structure equations with respect to the radial
variable  $\l$ along a  fixed spacelike autoparallel
$\g_{{}_{\pp_\l}}$ on $\S$.

At each step of the process one evaluates the (derivatives of)
structure forms at $\lambda=0$ thereby building up a Taylor series
in $\lambda$ for each. To initialise the process one must fix the
zeroth order terms in the expansion.
 The coframe field $\{ e^\a\}$ along $\g_{{}_{\pp_\l}}$ is defined to
 satisfy \be{\nabla_{\g^\prime{}_{{}_{\pp_\l}}} e^\a=0}
 i.e
 \be{i_{\pp_\l}\omega^\a{}_\b=0}
 in terms of the interior derivative of the connection 1-forms of
 $\nabla$ in the coframe $\OO$.
 Since all points of $\S$ can be connected by such autoparallels we
 conclude that $\omega^\a{}_\b$ must be independent of $d\,\l$ in
 the frame $\OO$ and in particular
 $\hat\omega^\a{}_\b(\hat\pp_j)=0$ on $\s$.  Since $\hat e ^i$ is
 dual to $\hX j =\hat\pp_j$ then $\hat e^i=\hat{dx^i}$. But
 $d\,x^i=p^i\, d\, \l +\l\, d\,p^i$ so $ e^i $ pulls back to
 $p^i\,d\l$ at $\l=0$. Also since $\dot\s=\frac{\hat\pp}{\pp t}$ then
 $\hat e^0= \hat{d\,t}$.
 In general \be{e^\a= e^\a(\pp_t)\, d\,t + e^\a(\pp_\l)\,d\,\l + e^\a(\frac{\pp}{\pp p^k})\, d\, p^k.}
 But along  $\g_{{}_{\pp_\l}}$,
 $ \pp_\l(e^\a(\pp_\l))=\nabla_{\pp_\l} (e^\a(\pp_\l)) =0 $,
since $e^\a$ is parallel and $\g_{\pp_\l}$ is an autoparallel.
Again, we may conclude that on $\S$ the structure function
$e^\a(\pp_\l)$ is independent of the coordinate $\l$. It follows
that $ e^i(\pp_\l)=e^i(\pp_\l)\vert_\s=p^i$ and
$e^0(\pp_\l)=e^0(\pp_\l)\vert_\s=\hat{dt}(\pp_\l)\vert_\s=0.$ Thus
one  may write:
 \be{e^\b = \ff\b \dd t + \Pi^\b \dd\l + \EE^\b}
where $\ff\b$ are 0 forms dependent on $t,\,\l,\,\ppp k$ and
\be{\Pi^0=0,\,\,\Pi^k=\ppp k}
 \be{\EE^\a = \delta^\a_0\,\YY +
\delta^\a_k\,\PHH k} with 1-forms  $\YY,\, \PHH k$  independent of
$\dd t,\, \dd\l$ but
 components dependent on $t,\,\l,\,\ppp k$.
Thus:
 \be{e^0=\ff 0 \,d\,t + \YY_j\,d\,p^j\label{E1}}
\be{e^k=\ff k\, d\,t + p^k\,d\,\l + \PHH k{}_j\, d \,p^j.
\label{E2}}
where \be{\hff 0 = 1\label{C22}} \be{\hYY = \hPHH k = \hff k =0.
\label{C11}}

Similarly, since $\omega^\a{}_\b(\pp_\l)=0$ on $\S$,
$\hat{\omega^\a{}_\b}(\hat{\pp}_j)=0$. Using $\frac{\pp p^j}{\pp
\l}=0$ and \be{ \frac{\pp}{\pp p^j}= \l\frac{\pp}{\pp x^j}- \l p^j
\sum_{k\ne j}\frac{1}{p^k}\frac{\pp}{\pp x^k}} one may write
\be{\omega^\a{}_\b = \AA\a\b \dd t + \CC\a\b\label{E3}} where
$\AA\a\b$ are 0 forms dependent on $t,\,\l,\,\ppp k$ and $\CC\a\b$
are 1-forms independent of $\dd t,\, \dd\l$ but with components
dependent on $t,\,\l,\,\ppp k$.

 Furthermore
\be{\omega^\a{}_\b(\pp_t)\vert_{\s(t)} = \hAA\a\b}
\be{\omega^\a{}_\b(\pp_{p^j})\vert_{\s(t)}\,\dd \ppp j = \hCC\a\b
= \omega^\a{}_\b(\pp_{x^j})\,\dd \ppp
j\vert_{\s(t)}=0.\label{ccval}}

The structure functions $\hAA\a\b$ can be determined in terms of
properties of the frame $\hat{\FF}$ on $\s$. Suppose the elements
of $\hat{\FF}$ satisfy:
 \begin{equation} F_{\dot\s} \hX \a=\Omega_\a{}^\b(\t) \hX
\b \label{FW}\end{equation}
 where $\Omega_{\a\b}=-\Omega_{\b\a}$, $\Omega_{0 j}=0$
and $\Omega_{i j}=\epsilon_{i j}{}^k\, \Omega_k(\t)$ in terms of
the alternating symbol $\epsilon^{ij k}$, indices on $\Omega$ and
$\epsilon$ being raised or lowered  with the Lorentzian metric
components $\eta=diag(-1,1,1,1)$. The (generalised) Fermi-Walker
derivative \CITE{perlick} above is defined by:

\be{F_{\dot\s} \hX \a= \nabla_{\dot\s}\hX \a + g(\dot\s, \hX \a)
\ACC_\s -g(\ACC_a,\hX \a)\dot\s} \be {=
\hat{\omega}^\mu{}_\a(\dot\s) \hX \mu + g(\dot\s, \hX \a) \ACC_\s
-g(\ACC_\s, \hX \a)\dot\s }

where $\ACC_\s \equiv \nabla_{\dot\s}\dot\s.$ Such a transport law
maintains $\hX 0$ tangent to $\s$ and $\hat{\FF}$ g-orthonormal
even though $\s$ need not be an autoparallel of $\nabla$. So given
$\dot\s=\hX 0\equiv \hat{\pp_t}$ with $\hat{e}^\nu(\hX
\mu)=\delta^\nu_\mu$ application of $\hat{e}^\nu$ to (\ref{FW})
yields: \be {\hat{\omega}^\nu{}_\a(\hat{\pp_t})=\Omega_\a{}^\nu -
\eta_{0\a}\,\hat{e}^\nu(\ACC_\s) + g(\ACC_\s,\hX\a)\, \delta^\nu_0
} or \be{\hat{\omega}^0{}_i(\hat{\pp_t})= g(\ACC_\s,\hX i)    }
\be{\hat{\omega}^i{}_j (\hat{\pp_t}  )=\Omega_i{}^j(\t)=\Omega_{i
j }(\t). } Thus \be{\hat{A}^\a{}_\b=
\hat{\omega}^\a{}_\b(\hat{\pp_t})\equiv
\Lambda^\a{}_\b\label{aaval}} say, is fixed in terms of the
orthornormal components of the {\it acceleration} $\ACC_\s$ of
$\s$ and the {\it instantaneous angular velocity} $\Omega_k$ of
the frame $\hat{\FF}$ along $\s$.  Since $\dot\s$ is normalised
and time-like the acceleration is orthogonal to $\dot\s$. It is
convenient to denote the acceleration by $\bfA$ and write \be{
\bfA_i=g(\ACC_\s,\hX i). } If one chooses $\Omega_k=0$ then
$\hat{\FF}$ is said to be a non-rotating frame along $\s$. Thus
with $\hCC\a\b=0$ and $\hAA\a\b=\L^\a{}_\b$ the equations
(\ref{E1}),
 (\ref{E2}), (\ref{E3}) subject to (\ref{ccval}), (\ref{aaval}),
(\ref{C11}), (\ref{C22}) can be substituted into the structure
equations \CITE{chern, sakai,RWT_book} that define the torsion and
curvature of $\nabla$:

 \be{
\dd e^\a = -\omega^\a{}_\b \wdg e^\b + T^\a \label{SE1} }
 \be{ \dd
\omega^\a{}_\b=-\omega^\a{}_\g\wdg \omega^\g{}_\b +
R^\a{}_\b\label{SE2} }

where $T^\a=\frac{1}{2}\,T^\a{}_{\mu\nu}\, e^\mu\wdg e^\nu$ and
$R^\a{}_\b=\frac{1}{2}\,\RR\a\b\mu\nu\,e^\mu\wdg e^\nu$ are the
torsion and curvature 2-forms in the frame $\OO$. Equations
\REF{SE1}, \REF{SE2} must be satisfied for all $\l$ along
auto-parallels that leave $\sigma(t)$ in any direction $\{p^j\}$
so by equating forms containing $\dd\l\wdg\dd t$ and $\dd\l\wdg\dd
p^k$ on each side one can find  differential equations for the
structure forms $\ff\a, \YY,\PHH i, \AA\a\b, \CC\a\b$:

\be{{\ff\a}{}^\prime = \AA\a k \ppp k + \TT\a k \mu\,\ff \mu \ppp
k} \be{ \YY^\prime=\CC 0 k \ppp k + \TT 0 k 0 \,\ppp k \YY + \TT 0
k n \ppp k \PHH n } \be{ {\PHH i}{}^\prime = \dd \ppp i + \CC i k
\ppp k + \TT i k 0 \, \ppp k \YY + \TT i k n \, \ppp k \PHH n }
\be{ {\AA\a\b}{}^\prime = \RR\a\b k\mu\,\ppp k \ff\mu} \be{ {\CC\a
\b}{}^\prime = \RR\a\b k 0 \,\ppp k \YY + \RR\a\b k n \, \ppp k
\PHH n .}

These have unique solutions satisfying the conditions:

\be{\hff 0 = 1} \be{\hYY = 0,\,\, \hPHH k = 0, \,\,\hff k = 0,\,\,
\hCC\a\b =0 } \be{ \hAA\a\b = \LL\a\b.}

By successively differentiating the differential equations above
with respect to $\l$ and evaluating the results at $\l=0$  one may
construct the Taylor series representation    of each analytic
structure form about $\l=0$ in terms of (derivatives of) the
components of the curvature and torsion tensor and the transport
functions $\L^\a{}_\b$ on $\s$. The first few iterations of this
process can be readily established. To first order in $\lambda$
one finds:

 \be{ \hff\a{}^\prime = \ppp k(\LL\a k  + \hTT\a k 0)}
 \be{\hYY{}^\prime = 0 }
 \be{\hPHH i{}^\prime = \dd \ppp i } \be{
\hAA\a\b{}^\prime = \hRR\a\b k 0\,\ppp k} \be{\hCC\a\b{}^\prime =
0}

Radial derivatives of the curvature and torsion start to appear at
the next order:

\be{\hff\a{}^\PP = \hRR\a k n 0 \ppp n\ppp k + \hTT\a k
0{}^\prime\,\ppp k + \hTT\a k \mu\, \ppp k \ppp n(\LL\mu n +
\hTT\mu n 0) } \be{ \hYY{}^\PP = \hTT 0 k n \ppp k \dd \ppp n  }
\be{ \hPHH i {}^\PP = \hTT i k n\, \ppp k \dd \ppp n  } \be{
\hAA\a\b{}^\PP = \hRR\a\b k 0{}^\prime \ppp k + \hRR\a\b k \mu
\ppp k \ppp n (\LL\mu n + \hTT\mu n 0) } \be{ \hCC\a\b{}^\PP =
\hRR\a\b k n \ppp k \dd \ppp n.  }

At the third order:

\begin{gather} \begin{split}\ \hff\a{}^\PPP =
&\hRR\a k n 0{}^\prime\,\ppp n\ppp k +  \hRR\a k n\mu\,\ppp n\ppp
k\ppp m(\LL\mu m +\hTT\mu m 0) +
\\ &  + \hTT\a k 0{}^\PP\,\ppp k + 2\,\hTT\a k\mu{}^\prime\,\ppp
k\ppp n (\LL\mu n + \hTT\mu n o) +
\\ & + \hTT\a k\mu\,\hRR\mu m n 0\,\ppp k\ppp n \ppp m +\hTT\a
k\mu\,\hTT\mu m 0{}^\prime \,\ppp k\ppp m +
\\ & +  \hTT\a k\mu\,\hTT\mu m \nu\, \ppp k\ppp m \ppp m (\LL\nu n
+ \hTT\nu n 0 )
\end{split}
\end{gather}

\be{\hYY{}^\PPP = \hRR 0 k m n\,\ppp m\ppp k \dd\ppp n + 2\hTT 0 k
n {}^\prime\,\ppp k \dd\ppp n +\hTT 0 k \gamma\, \hTT\gamma n q
\ppp k \ppp n\dd\ppp q} \be{\hPHH i{}^\PPP = \hRR i k n q \ppp n
\ppp k \dd\ppp q + 2 \hTT i k n{}^\prime\,\ppp k\dd\ppp n + \hTT i
k \gamma \,\hTT\gamma n q  \ppp k \ppp n\dd\ppp q }
\begin{gather} \begin{split}\  \hAA\a\b{}^\PPP =
& \hRR\a\b k 0 {}^\PP\,\ppp k +2\,\hRR\a\b k \mu{}^\prime\,\ppp
k\ppp n(\LL\mu n + \hTT\mu n 0) +
\\ & + \hRR\a\b k\mu\,\hRR\mu n q 0 \ppp k\ppp n\ppp q + \hRR\a\b k
\mu\,\hTT\mu n 0{}^\prime \ppp k\ppp n +
\\ & +  \hRR\a\b k \mu \, \hTT\mu m \nu\,\ppp k\ppp m\ppp n (\LL\nu
n + \hTT\nu n 0)
\end{split}
\end{gather}
\be{ \hCC\a\b{}^\PPP =
 2\, \hRR\a\b k n {}^\prime\,\ppp k\dd \ppp n +
\hRR\a\b k \delta\,\hTT \delta m q \ppp k\ppp m \dd\ppp q. }

In these expressions the directional derivatives $\partial^m{}_\l$
for the appropriate order $m$ are to be expressed in terms of
derivatives with respect to $x^k$:
\be{\partial^m{}_\l=p^{i_1}\,p^{i_2}\ldots
p^{i_m}\,\pp_{i_1}\pp_{i_2}\ldots \pp_{i_m}} and the result of
applying $ \partial^m{}_\l $ evaluated on $\s(t)$. The metric
tensor  \be{g=-e^0\otimes e^0 + e^k\otimes e^k \label{metric}} is
now given in terms of the above as a Taylor series to order
$\l^3$, and generalises the results of \CITE{MM, ni, lini,
marzlin}.

\section{Identities from Gauge Covariance }

Attention is restricted to background gravitational and
electromagnetic fields. The former is described in terms of a
metric tensor $g$ and a metric-compatible connection $\nabla$. The
latter is given in terms of a 2-form $F$. The basic gravitational
variables will be a class of {\it arbitrary} local orthonormal
1-form coframes $\{e^\a\}$ on spacetime related by $SO(1,3)$
transformations. In such frames the connection $\nabla$ is
represented by the spacetime 1-forms $\{ \omega^\a{}_\b\}$.
 The electromagnetic
$U(1)$ connection can be represented locally by a spacetime 1-form
$A$ and $F=\dd A$. All other (matter) fields are denoted
collectively by $\mat$. It is assumed that in a background
electromagnetic and Einstein-Cartan gravitational field described
by $\{e,\omega\}$ the matter field interactions can be derived
from some action functional: \be{S[e,\omega,A,\mat]\equiv
\int_M\L} where the 4-form $\L$ is constructed from the field
variables $\{e,\omega,A,\Phi\}$ and their derivatives, so that $S$
is invariant under change of local $U(1)$ and Lorentz group
transformations {\footnote{In theories with gravitational scalar
fields it is possible to extend the local gauge symmetries to
include Weyl scale covariance and an associated connection
1-form.}}. Its invariances impose restrictions on the variational
derivatives that arise in the expression \be{\int_M \LX\L= \int_M
(\t_\mu\wdg \LX e^\mu + S_\mu{}^\nu\wdg\LX \omega^\mu{}_\nu + j
\wdg \LX A + \EE \wdg \LX \mat) \label{var}}
 where the vector field $X$ generates a
local spacetime diffeomorphism and $\LX$ denotes the Lie
derivative with respect to $X$. Since $\LX\L=i_X\dd\L+\dd i_X\L$
and $\dd\L=0$ the left hand side of (\ref{var}) is zero if $X$ has
compact support.  The 3-forms $\t_\mu, S_\mu{}^\nu,$ and $j$ are
the basic variables in our  balance laws and may be termed the
source currents for gravitation and electromagnetism.  In the
absence of electrically charged matter and a background
electromagnetic field a recent derivation of the identities that
arise from gauge covariance can be found in \CITE{imb}. The
inclusion of electromagnetic interactions demands only a minor
modification to these arguments but gives rise to additional
terms. A sketch of the arguments follows.

Since the matter field equations are to be satisfied, $\EE=0$.
Using the identity
 \be{\LX A=i_X\dd A+\dd i_X A}
  and recognising
that under a $U(1)$ gauge transformation $A\mapsto A +\dd f$ for
some spacetime scalar field $f$ one may write
 \be{\LX A =i_X F
+\gg{U(1)} A}
 in terms of the generator $\gg{U(1)}$ of a $U(1)$
gauge transformation. In a similar way with the aid of the
structure equations \REF{SE1}, \REF{SE2} the terms $\LX e^\mu$ and
$\LX \omega^\mu{}_\nu$ can be written in terms of the generators
of local Lorentz transformations and exterior covariant
derivatives \CITE{RWT_book}: \be{ \int_M \LX\L= \int_M( \t_\mu\wdg
i_X T^\mu + \t_\mu \wdg D(i_X e^\mu ) + S_\mu{}^\nu \wdg i_X
R^\mu{}_\nu  } \be{ + j\wdg i_X F + j \wdg \gg {U(1)}A - \t_\mu
\wdg \gg {SO(3,1)} e^\mu - S_\mu{}^\nu \wdg \gg {SO(1,3)}
\omega^\mu{}_\nu ).\label{var2}}

By gauge invariance of the action $\int_M\Lambda$
\be{\int_M(\gg{U(1)}+ \gg{SO(1,3)})\L=0 } so the last three terms
in (\ref{var2}) are zero. If the arbitrary components $i_X e^\mu$
of $X$ have compact support one may replace the second term in
(\ref{var2}) by $D\t_\mu\wdg i_X e^\mu$ and conclude: \be{\int_M(
D\t_\a+\t_\mu\wdg i_{X_\a} T^\mu + S_\mu{}^\nu\wdg i_{X_\a}
R^\mu{}_\nu + j\wdg i_{X_\a} F)\,\nn^\alpha=0\label{BL11}} where
$\{X_\a\}$ is the dual frame and $\{\nn^\mu\}$ a set of test
0-forms. Similarly invariance of the action $\int_M\L$ under
$\gg{SO(1,3)}$ alone gives: \be{\int_M( D S_\mu{}^\nu-
\frac{1}{2}(\t_\mu\wdg e^\nu - \t^\nu\wdg e_\mu))\,\nn^\mu{}_\nu=0
\label{BL22}} for  a set of test 0-forms $\{\nn^\mu_\nu\}$. For
smooth tensors on a regular domain of $M$ one deduces the local
identities:

\be{D\t_\a+\t_\mu\wdg i_{X_\a} T^\mu + S_\mu{}^\nu\wdg i_{X_\a}
R^\mu{}_\nu + j\wdg i_{X_\a} F = 0\label{BL1}}

\be{ D S_\mu{}^\nu= \frac{1}{2}(\t_\mu\wdg e^\nu - \t^\nu\wdg
e_\mu).\label{BL2}}

In these expressions
$$ D\,\t^\a\equiv \dd \t^\a + \omega^\a{}_\b\wdg \t^\b$$

$$ D\,S^{\a\b}\equiv \dd S^{a\b} + \omega^\a{}_\g \wdg S^{\g\b} +
\omega^\b{}_\g \wdg S^{\a\g}.$$

Finally invariance of the action under $\gg{U(1)}$ alone yields:
\be{\int_M\gg{U(1)}\L = \int_M j\wdg \gg{U(1)} A=\int_M j\wdg \dd
f = \int_M f \dd j=0. } For smooth $j$ on a regular domain \be{
\dd j=0 \label{BL3}} since $f$, with compact support, is
arbitrary.

It should be stressed that (\ref{BL1}), (\ref{BL2}), (\ref{BL3})
are {\it identities} in a field theory satisfying the conditions
used in their derivation. However in a dynamical scheme that
attempts to approximate the currents $\t_\a, S_\mu{}^\nu, j$ they
provide powerful constraints. Thus  (\ref{BL3}) is clearly a
conservation law (in an arbitrary spacetime background), since for
any 4 dimensional domain $\SU\subset M$ with
$\pp\SU=\Sigma_1-\Sigma_2 + \CCC$ and $\Sigma_1,\Sigma_2$ disjoint
spacelike hypersurfaces in $M$, and $j$ regular on $\SU$:
\be{0=\int_\SU\dd j =\int_{\pp\SU}\,j = \int_{\Sigma_1} j -
\int_{\Sigma_2} j  } if $\int_\CCC j =0$. The integral
$\int_{\Sigma_1} j $ is the electric charge content of $j$ in
$\Sigma_1$.

\section{Models for the Source Currents}

The balance laws (\ref{BL1}), (\ref{BL2}), (\ref{BL3}) above are
devoid of predictive content without further information that
relates the source currents to each other and the background
fields. Such {\it equations of state} serve to define the
dynamical variables in the theory and reduce the number of degrees
of freedom subject to temporal evolution. The established laws of
Newtonian and special relativistic field-particle interactions
offer valuable guides  since they should be recovered in
appropriate limits. Indeed an effective source model should enable
one to recover the framework of non-relativistic continuum
mechanics from the balance laws in a Minkowski spacetime free of
curvature and torsion. More generally it is known how to model a
spinless pressureless fluid in a pseudo-Riemannian spacetime with
curvature and recover the motion of dust as geodesics associated
with the Levi-Civita connection \CITE{ehlers:1971, fluid1}. In
spaces with curvature and torsion analogous equations for spinning
hyperfluids can be constructed \CITE{TrautI, TrautII, TrautIII,
fluid3, fluid4, fluid2}.

In \CITE{dixon4:1974}  Dixon  considered the possibility of using
the pseudo-Riemannian balance laws to construct a multipole
analysis of the dynamics of an extended relativistic structure.
Following work by Pryce \CITE{pryce} he gave cogent arguments for
a type of subsidiary condition that eliminated peculiar motions of
electrically neutral mass monopoles in the absence of gravitation.
The approach advocated here is based on a similar  description of
a particle in terms of the evolution of appropriate multipoles
along the timelike worldline $\s$. The structure of the particle
is therefore given in terms of components of the source currents
in the Fermi frame $\FF$. Thus we seek a consistent dynamical
scheme that will determine these components and the history $\s$
where the balance laws are satisfied to some order in a suitable
multipole expansion.

The basic assumption is that the source currents for a classical
particle are localised on the tubular region $\UU$ about $\s$.
Thus if $\Sigma_t\subset\UU$ is the hypersurface $t=\,$constant
then the test functions in (\ref{BL1}), (\ref{BL2}), (\ref{BL3})
are rapidly decreasing as a function of the {\it geodesic
distance} $\l$ in the 3-ball $\Sigma_t$ centred on
$\s(t)$\footnote{ To lowest non-trivial order the calculations
below are insensitive to the precise fall off with $\lambda$ }.
The precise structure of the stress-energy 3-form currents $\t^\a$
defines the nature of "particle" variables used to pass from a
field to a particle description. The corresponding choice of
stress-energy tensor field $T=\star^{-1}\t_\a \otimes e^\a$ is
analogous to a choice of constitutive relation in continuum
mechanics relating configuration stresses and powers to strains
and work. The operator $\star$ here denotes the Hodge
map\footnote{In terms of an orthonormal basis the volume form
$\star 1$ is chosen as $e^0\wdg e^1\wdg e^2\wdg e^3$. } with
respect to the metric $g$. The choice should naturally have the
proper Newtonian limit for a spinless particle. We shall also
ensure that the Matthisson-Papapetrou equations for electrically
neutral spinning particles emerge in a torsion-free
pseudo-Riemannian spacetimes \CITE{pap, Tul}.  Thus in the absence
of self-fields (see later) consider:

\be{\t^\a=P^\a\se 0 + \chi F^{\a\b}\,\SS_{\b\mu}\se
\mu\label{taucurrent}} \be{S_\mu{}^\nu=\SS{}_\mu{}^\nu
\se0\label{spincurrent}} \be{j=\rho\se 0.\label{emcurrent}}

in terms of the scalar components $P^\a$,
$\Sigma^{\a\b}$,$F^{\a\b}$ $\rho$ on $\UU$ and constant $\chi$.

Since the background fields are assumed regular on $\s$ the
components of the curvature, torsion and $F$ as well as $P^\a$,
$\Sigma^{\a\b}$, $\rho$ can each be expanded as a Taylor series of
the form (13). The coframe $e^\a$ and connection forms
$\omega^\a{}_\b$ are also available as expansions in $\l$ so the
coefficient of the test function in the integrand of each balance
can  be expressed as a power series in $\l$ with a multinomial
dependence on $p^1,p^2,p^3$. To extract dynamic information one
now truncates each series to some order and integrates over
$\Sigma_t$. This may be achieved by writing
$$p^1=\sin\theta\cos\phi,\quad p^2=\sin\theta\sin\phi,\quad p^3=\cos\theta$$
and noting that if  the positive integers $n_1,n_2,n_3$ are all
even then

\be{\int_{S^2}(p^1)^{n_1} \, (p^2)^{n_2} \,(p^3)^{n_3}
\,d\Omega=\frac{1}{2} \frac{\Gamma(\frac{n_1+1}{2})
\Gamma(\frac{n_2+1}{2}) \Gamma(\frac{n_3+1}{2})  }
{\Gamma(\frac{n_1+n_2+n_3}{2})}.}

If any integer is odd then the integral is zero. Since

$$ e^0\wdg e^1 \wdg e^2 \wdg e^3 = \lambda^2 \sin\theta \dd\theta
\wdg\dd \phi\wdg \dd\lambda\wdg\dd(ct) + O(\lambda^3)$$

the lowest order of relevance includes terms up to order $\l^2$.
The resulting equations yield ordinary differential equations for
the dynamical variables $\hat\Sigma^{\a\b}, \hat P^\a$ since the
conservation equation $\dd j=0$ requires $q=\int_{\Sigma_t}j$ to
be a constant. It is convenient to introduce new variables:
$$P^i=\hat P^i$$
$$P^0=\hat P^0$$
$$s_i=\frac{1}{2}\epsilon_{i j k}\,\hat\Sigma^{j k}$$
$$ h_k=\hat\Sigma^{0 k}$$
$$ B_i = \frac{1}{2} \epsilon_{i j k}\, \hat F^{j k}$$
$$ E_i = {\hat F}^{0}{}_i$$
and denote ordinary derivatives with respect to $c\,t$ by an
over-dot.

In terms of the spin 2-form
$$\SSS\equiv\frac{\Sigma_{\a\b}}{2}\,e^\a\wdg e^\b$$ and the form
$\tilde V=e^0$ one has $i_V\star \SSS=\star(\tilde V \wdg \SSS)$.
Since $V$ restricts to the tangent to $\s$ one readily recognises
that $s=s_k\,e^k=-i_V\,\star\SSS$ is dual to the Pauli-Lubanski
spin vector on $\s$ where $\hat\SSS=h\wdg e^0 + i_V\star s$ and
the 1-forms $h=h_k\,e^k$ and $s$ are spatial:
$i_{X_0}\,h=i_{X_0}\,s=0$. Similarly with $\hat P\equiv
P^0\,X_0+P^k\,X_k$ one has on $\s$: \be{ i_{\hat P}\hat\SSS=(h_k
P_k) \,e^0-P^0\,h + \epsilon_{i j k}\,s^i\,p^j\,e^k }

Then with respect to a non-rotating Fermi frame ($\Omega_k=0$),
the equations (\ref{BL1}), (\ref{BL2}), (\ref{BL3}) give:

\be{\dot\bfs=\bfh\times \bfA + \frac{\chi}{2}(\bfB\times \bfs +
\bfE\times \bfh)\label{heqn}} \be{\dot\bfh=\bfA\times \bfs -
\frac{1}{2}\bfP + \frac{\chi}{2}(\bfB\times \bfh + \bfs\times
\bfE)\label{seqn}}
\begin{gather} \begin{split}\dot P^0 =
&  -\bfA\cdot\bfP + \chi[(\bfh\cdot\bfE)^{\dot{}} +
\bfA\cdot(\bfB\times \bfh) + \hTT 0 0 k (\bfs\times\bfE)_k  +  \\
& + \hat B_i\,s_j\,\hTT j 0 i - \hTT j 0 j (\bfB\cdot\bfs)- \hat
E_j\,\hat h_k\,\hTT j 0 k ] \label{p0eqn}
\end{split}
\end{gather}

\begin{gather} \begin{split} \left(\dot\bfP\right)_k =
& \hat R_{\a\b k 0}\,\Sigma^{\a\b}-\hTT\b 0 k \,P_\b
-(\bfA)_k\,P^0 -q(\bfE)_k +\\ & +\chi[ (\bfB\times\bfh)_k^{\dot{}}
+(\bfA)_k(\bfE\cdot\bfh)
%
 +\\ & + \hat E_i\,h^j\, \hTT i k j -
(\bfE\times \bfs)_j\,\hTT 0 j k  + (\bfB \times\bfh)_j\,\hTT j 0 k
- (\bfE\cdot\bfh)\,\hTT 0 0 k\\ & -\hTT j k n \,\hat B_n\,s_j +
\hTT j k j \,(\bfB \cdot \bfs)].\label{pkeqn}
\end{split}
\end{gather}

\section{Involution and Constitutive Relations}

For given external fields $\bfE, \bfB, \hTT\a \b\mu,
\hRR\a\b\mu\nu$ equations (\ref{heqn}), (\ref{seqn}),
(\ref{p0eqn}), (\ref{pkeqn})  are 10 ordinary differential
equations involving 13 functions ($\bfs,\bfh,\bfA,P^\a$) of $t$
and the constants $\chi,q$. Additional information is required to
render these equations useful. Since radiation reactions due to
{\it self fields} have been neglected such information should
provide a means of determining the acceleration $\bfA$ of the
curve $\s$ in terms of other variables, that may themselves
satisfy a differential system. Since
$\bfA_k=(\nabla_{\dot\s}\dot\s)_k$, the equation for $\s$ can then
be determined in any coordinate system. For consistency the final
set of equations should reduce to  an involutive differential
system. So, suppose $\xi_A(t)$ denotes collectively the variables
$(\bfs,\bfh,P^\a)$ {\it and} the background fields whose $t$
dependence on $\s$ is supposed known. Write (\ref{seqn}),
(\ref{heqn}), (\ref{p0eqn}), (\ref{pkeqn}) and the time
derivatives of the background field components as \be{
\dot\xi_A=W_A(\xi,\bfA,\chi,q).\label{sys} } Let \be{
\KK_B(\xi,q,\chi)=0\label{subs1}} be a set of scalar subsidiary
conditions (constitutive relations indexed by $B$) that is
appended to the above system. Then the requirement $ \dot\KK_B=0$
implies \be{\frac{\partial\KK_b(\xi,q,\chi)}{\partial \xi_A}
W_A(\xi,\bfA,\chi,q)\vert_{\KK}=0 \label{subs1A}} where the
relation (\ref{subs1}) and implicit $t$ dependence of the
background has been used.  If, for any $\chi,q,$ (\ref{subs1A})
determines $\bfA=\bfA(\xi,q,\chi)$ one may insert this in
(\ref{sys}) and solve for $\xi_A$ given initial conditions. Such a
system then determines the acceleration $\bfA$. However
(\ref{subs1A}) may only determine some of the components of
$\bfA$, or the solution may depend on the nature of the external
fields or the values of the parameters $\chi,q$.  If no suitable
choice can be found for generic external fields then $\KK_B=0$ is
an unsuitable condition for the source model under consideration
and a better condition (or model) is required. Another possibility
is that (\ref{subs1A}) leads to  new conditions, say
$L_{B^\prime}(\xi,\chi,q)=0$ independent of $\bfA$. In this case
$\dot L_{B^\prime}=0$ requires: \be{\frac{\partial
L_{B^\prime}}{\partial \xi_A}\,
W_A(\xi,\bfA,\chi,q)\vert_{L,\KK}=0 } These equations are now
analysed in the same way as (\ref{subs1A}) and the procedure
repeated until $\bfA$ is determined for a consistent $\chi,q$, or
the process exhibits a manifest inconsistency.

For models with more parameters and more components in the source
currents the strategy is similar. Finding a constitutive relation
that renders a differential-algebraic-system involutive is, in
general, non-trivial and further recourse to weak field
approximations may be necessary to make progress. The above
strategy is based on the premiss that {\it radiative} or
self-field contributions to the balance laws are not relevant.
Such contributions are expected to yield terms containing higher
derivatives of $\bfA$ (that may be representations of a
differential-delay system) and the above involution analysis must
be generalised accordingly. The inclusion of such terms is a
challenging and important issue that will not be pursued further
here.

To illustrate this procedure equations  (\ref{heqn}),
(\ref{seqn}), (\ref{p0eqn}), (\ref{pkeqn}) will be supplemented
with the relations: \be{ \left (\t_\a  -
 \chi F_\a{}^\b\,\Sigma_{\b\mu}\,\star e^\mu  \right ) \wdg \star S^{\a\delta}=0.}
 With the source model above these become the Tulczyjew-Dixon conditions:
\be{P_\a\,\Sigma^{\a\b}=0} or \be{i_{\hat
P}\,\hat\SSS=0\label{DIX}} in terms of the spin 2-form. In terms
of the particle variables on $\s$, \REF {DIX}  gives:
\be{\bfh\,P^0=\bfs\times \bfP\label{DIX1}} and this implies
\be{\bfP \cdot \bfh =0} \be{\bfs\cdot\bfh=0.} By contrast the
condition \be{i_V\,\hat\SSS=0\label{MPcnd}} gives $\bfh=0$ .

 An immediate consequence of \REF {DIX} from
 (\ref{heqn}), (\ref{seqn}) is:
  \be{\left(\hat\Sigma_{\a\b}{\hat{\Sigma^{\a\b}}}\right )
^{\dot{}}=(\bfs\cdot\bfs -\bfh\cdot\bfh)^{\dot{}}=0} in any
background.

\section{Special Cases}

It is of interest to analyse (\ref{heqn}),  (\ref{seqn}),
(\ref{p0eqn}), (\ref{pkeqn}) in Minkowski spacetime devoid of
curvature and torsion. First consider the case where $\chi=0$.
Then \be{ \dot\bfs=\CROSS \bfh \bfA \label{flatseqn}} \be{
\dot\bfh=\CROSS \bfA \bfs -\frac{1}{2}\,\bfP\label{flatheqn}}
\be{\dot P^0=-\DOT\bfP\bfA\label{flatp0eqn}}
\be{\dot\bfP=-P^0\bfA-q\,\bfE\label{flatpeqn}.} These immediately
imply: \be{(\DOT\bfs\bfs-\DOT\bfh\bfh)^{\dot{}}=\DOT\bfh\bfP}
\be{(-(P^0)^2+\DOT\bfP\bfP)^{\dot{}}=-2q\,\DOT\bfP\bfE.} If the
spin current is zero, $\bfs=0$ and $\bfh=0$ (so
$\Sigma_{\a\b}\,\Sigma^{\a\b}=0$), then the above equations reduce
to \be{\bfP=0}\be{\dot P^0=0} and \be{\bfA=-q\,\frac{\bfE}{P^0}}
without further conditions. Introducing the constant of motion
$P^0=m$ the covariant equation of motion for the particle follows
from the last equation:\be{\nabla_{\dot\s}\dot\s=\frac{q}{m}
\widetilde{i_{\dot\s}F}\label{LFeqn}} in terms of the Levi-Civita
connection $\nabla$ with $\tilde{\a}=g^{-1}(\a,-)$ for any 1 form
$\a$. One recognises the equation of motion of a spinless particle
with mass $m$ and electric charge $q$ \CITE{swu:1977}. If $F=0$
the particle follows a geodesic of this $\nabla$.

Suppose next that $\Sigma_{\a\b}\,\Sigma^{\a\b}\ne 0$. The early
descriptions of spinning point particles (see also \CITE{mat,pap})
employed  conditions leading to $\bfh=0$ or $i_V\,\SSS=0$. With
the equations \REF{flatseqn}, \REF{flatheqn}, \REF{flatp0eqn},
\REF{flatpeqn} this yields:
\be{\dot\bfs=0}\be{\bfP=2\CROSS\bfA\bfs\label{PAS}}\be{\dot
P^0=0}\be{\dot\bfP=-P^0\,\bfA-q\,\bfE.} Then
\be{(\DOT\bfs\bfs)^{\dot{}}=0}\be{({P_\a\,P^\a})^{\dot{}}=-2q\,\DOT\bfP\bfE}
and $\bfs=\bfs_0\ne 0$ and $P^0=m\ne 0$ are constants of the
motion. Differentiating \REF{PAS} gives \be{2\dot\bfA \times
\bfs_0= -m\bfA -q\,\bfE\label{PASS}.} If $q=0$ this implies
$\DOT\bfA\bfs_0=0$ and hence $\DOT{\dot\bfA}\bfs_0=0$. Taking the
cross product of \REF{PASS} (with $q=0$) with $\bfs_0$ then gives:
\be{\dot\bfA={\mathbf\Omega_0} \times \bfA} where
${\mathbf\Omega_0}=-\frac{m}{2\modsq {\bfs_0}}\bfs_0$. Thus, in
addition to geodesic motion given by $\bfA=0$, if $\bfA(t_0)\ne 0$
for any $t_0$, solutions exist where the acceleration exhibits  a
rotation with angular velocity ${\mathbf\Omega_0}$:
\be{\bfA(t)={\cal R}(t)\,\bfA(0)} where ${\cal R}$ is a time
dependent $SO(3)$ matrix describing a rotation with angular
velocity ${\mathbf\Omega_0}$. Such a motion is unnatural for a
classical particle free of electromagnetic and gravitational
interactions. The existence of such solutions is one of the
reasons that the Tulczyjew-Dixon condition is to be preferred over
\REF{MPcnd}. If $q\ne0$ then the additional terms yield the
modified equation \be{\dot\bfA=\CROSS{{\mathbf\Omega_0}}{\bfA} -
\frac{q}{m}\CROSS{\bfE}{\mathbf\Omega_0} -4\frac{q}{m}
{\mathbf\Omega_0}\,(\DOT{\dot\bfE}{\mathbf\Omega_0})\,\left(
\frac{\modsq{\bfs_0}}{m^2} \right)} which also exhibits unnatural
motions requiring a specification of an initial acceleration.

With \REF{DIX} instead of \REF{MPcnd}  equations \REF{flatseqn},
\REF{flatheqn}, \REF{flatp0eqn}, \REF{flatpeqn} reduce to a system
that determines $\bfA$ rather than $\dot\bfA$ and hence fixes the
motion of the particle in terms of its initial position and
velocity. Differentiating the condition \REF{DIX1} and using
equations \REF{flatseqn}, \REF{flatheqn}, \REF{flatp0eqn},
\REF{flatpeqn} yields a new (vector) condition:
\be{P^0\bfP=2q\CROSS\bfs\bfE\label{newcnd}} and hence
$\DOT\bfs\bfP=0$ and $\DOT\bfE\bfP=0$. The invariant $P_\a\,P^\a$
is a constant of the motion which we take to be the non-zero
constant $-m^2$. Differentiating \REF{newcnd} and using the
previously obtained equations gives rise to the equation: \be{
(P^0)^4\bfA+(P^0)^3\,q\,\bfE+2q\,(P^0)^2\,
\CROSS\bfs{\dot\bfE}-4q^2\,(\DOT\bfA\bfs)\,\CROSS\bfE\bfr=0\label{Aeqn}
} where $\bfr\equiv \CROSS\bfs\bfE$. For a generic $\bfE$,
($\bfr\ne 0$), one may take $\bfr, \bfs, \bfE$ as a basis in which
to evaluate $\bfA$ from \REF{Aeqn} in terms of
$P^0,\bfs,\bfE,\dot\bfE$. Thus \be{\bfA=A_r\,\bfr + A_s\,\bfs +
A_E\,\bfE \label{ACCEL}} where \be{ A_r=\frac{2q}{\modsq\bfr\,
{P^0}^2}\,\DOT\bfr{(\CROSS{\dot\bfE}{\bfs})} } \be{
A_E=\frac{\modsq{\bfs}}{\modsq{\bfr}}\,\DOT\bfA\bfE -
\frac{(\DOT\bfE\bfs)}{\modsq{\bfr}}  \DOT\bfA\bfs }
\be{\bfA_s=-\frac{(\DOT\bfE\bfs)}{\modsq{\bfr}}\,\DOT\bfA\bfE  +
\frac{{\modsq{\bfE}}}{\modsq{\bfr}} \, \DOT\bfA\bfs     } and \be{
\DOT\bfA\bfE= -\frac{q\modsq{\bfE}}{(P^0)}   -
\frac{2q}{(P^0)^2}\, \DOT{\bfE}{(\CROSS\bfs{\dot\bfE})} }
\be{\DOT\bfA\bfs= -\frac{q P^0(\DOT\bfE\bfs)} {\left( (P^0)^2 +
2q\, (\DOT\bfE\bfh) \right)} }

with

\be{ \bfh=\frac{2q}{(P^0)^2}\,\CROSS\bfs{(\CROSS\bfs\bfE)}   .}
 The above expression for $\bfA(P^0,\bfs,\bfE,\dot\bfE)$ can now
 be inserted into \REF{flatseqn}, \REF{flatp0eqn} yielding:
\be{ \dot P^0 =
-\frac{2q}{P^0}\,\DOT\bfE{(\CROSS\bfA\bfs)}\label{pop1}}
\be{\dot\bfs=  -\frac{2q}{(P^0)^2}
\CROSS\bfA{(\CROSS\bfs{(\CROSS\bfs\bfE)})}\label{pop2} . }

Given initial values for $\bfs$ and $P^0$  solutions of the
coupled equations \REF{pop1} and \REF{pop2} can be used to
determine the acceleration from \REF{ACCEL}. Since $\bfh=0$ and
$\dot\bfs=0$ if $\bfE=0$ we set $\bfs\vert_{\bfE=0}=\bfs_0$
constant. Thus the constant of motion $\DOT\bfs\bfs - \DOT\bfh\bfh
= \modsq{{\bfs_0}}$ identifies $\bfs_0$ as the classical spin of
the particle in the absence of electromagnetic interactions.
Similarly $m$ may be identified with the particle's rest mass in
the absence of such interactions. Thus at any $t=t_0$:
\be{{({P^0}(t_0))}^2 = m^2 + \frac{4q^2}{{{P^0}(t_0)}^2} \,
\modsq{{\CROSS\bfs\bfE}}(t_0) }

\be{\modsq{{\bfs(t_0)}}=   \modsq{{\bfs_0}}  +
\frac{4q^2}{{{P^0}(t_0)}^4} \, \modsq{    {({
\CROSS\bfs{(\CROSS\bfs\bfE)} })} } (t_0). }

If the particle moves slowly with velocity ${\bf v}$ in an
inertial Minkowski frame, where the electromagnetic field has
components ${\bf {{\cal E}=0, \cal{B}}}$, \CITE{splitem}, then in
the Fermi frame of the particle $\bfE\simeq \frac{{\bf v}}{c}
\times {\bf{\cal B}},\,\, \bfB\simeq {\bf{\cal B}}$. Thus the spin
coupling to the electromagnetic field indicates that it does not
have a magnetic moment. Such an interaction arises from the
inclusion of the terms in the current that depend on $\chi$. To
see this consider $\chi<<1$. Then the equations \REF{flatseqn},
\REF{flatheqn}, \REF{flatp0eqn}, \REF{flatpeqn} are replaced by:

\be{\dot\bfh = -\frac{\bfP}{2} + \CROSS\bfA\bfs - \frac{\chi}{2}
\left( \CROSS\bfE\bfs + \CROSS\bfh\bfB \right)  } \be{ \dot\bfs =
\bfh \times \bfA + \frac{\chi}{2} \left( \CROSS\bfB\bfs +
\CROSS\bfE\bfh \right)  } \be{ \dot P^0 = -\DOT\bfP\bfA +
\chi\,\left( \DOT\bfE{(\CROSS\bfA\bfs)} -\frac{1}{2}\,\DOT\bfP\bfE
\right) + O(\chi^2) } \be{\dot\bfP = -q\,\bfE -P^0\bfA + \chi
\left( \CROSS\bfB{(\CROSS\bfA\bfs)} + \bfA\,(\DOT \bfE\bfh) -
\frac{1}{2}\, \CROSS\bfB\bfP \right) + O(\chi^2). }

Consider static fields ($\dot\bfE=0,\,\,\dot\bfB=0$). Then after
differentiating the condition \REF{DIX},  taking scalar products
of the result with $\bfs$ and $\bfP$ and using the above equations
along with the relations $\DOT\bfP\bfs=0$ and $\DOT\bfs\bfh=0$ an
additional constraint arises that for $P^0\ne 0$ requires $\bfP=0$
at this order. This in turn implies (for $\bfs\ne0$) $\bfh=0$ and
hence $P^0=m$ constant. Furthermore now: \be{\bfA =
-\frac{q}{m}\,\bfE}\be{\dot\bfs = \frac{\chi}{2}\,\CROSS\bfB\bfs.
} But since $\dot\bfh = 0$,
$\CROSS\bfA\bfs=\chi\,(\CROSS\bfE\bfs)/2$ or \be{\chi =
-\frac{2q}{m}\label{chival}.} Thus, in this small $\chi$,
stationary field, limit one has \REF{LFeqn} and
\be{\dot{\hat\SSS}_{\a\b} =
\frac{1}{2}\,\hat{F}_{\a\mu}\hat\SSS^\mu{}_\b} and one may
identify the interaction of electromagnetism with a charged
particle possessing a magnetic moment related to its spin with a
gyromagnetic ratio $2$ \CITE{holten}.
 Thus the source currents \REF{taucurrent}, \REF{spincurrent}, \REF{emcurrent},
  are appropriate for the
 description of a spinning particle of mass $m$, electric charge
 $q$ and gyromagnetic coupling defined by \REF{chival}.

\section{Generalisations}

\def\s{\Sigma}

A relativistic  description of a material rod involves a
2-dimensional immersed timelike submanifold $\s$ of spacetime (the
rod \emph{worldsheet}) with attached material body frames that
relate dilation,
 bending, shear and torsional deformations to the immersion. As before, to obtain
such an immersion from an interacting field description requires a
reduction, or truncation, of  field degrees of freedom. This is
most naturally accomplished in terms of a field of local adapted
ortho-normal spacetime frames  generated by the transport of local
material body frames along spacelike autoparallels of the Lorentz
connection $\nabla$ to points in a neighbourhood  $\UU$ of  the
worldsheet $\s$. Dimensional reduction of the stress-energy and
spin Bianchi identities, and entropy imbalance relations on $\UU$,
is performed in new Fermi coordinates $\{
\sigma_0,\sigma_1,\lambda,p_1,p_2\}$ (with $p_1^2+p_2^2=1$)
adapted to such body frames. Their definition in terms of the
exponential maps is a generalisation of that given earlier with
$\s$ replacing the worldline $\sigma$.  In such coordinates the
history of the rod $\s$ is given as the timelike submanifold
$\{\lambda p_1=0,\lambda p_2=0\}$ for some range of
$\{\sigma_0,\sigma_1\}$ and the local coframe on $\UU$ may be
written: \be{e^j=f^j{}_i\,d\sigma^i + \phi^j}
 \be{e^\b=f^\b{}_i\,d\sigma^i+p^\b\,d\lambda +\phi^\b}
 for $i=0,\,1$ and $\a,\,\b=2,3$,
where $f^j{}_i,\,f^\b{}_i$ are 0 forms dependent on
$\sigma^i,\,\l,\,p^\a$  and with 1-forms  $\phi^\aa$
($\aa=0,1,2,3$) independent of $d\sigma^i,\, \dd\l$ but
 components dependent on $\sigma^i,\,\l,\,p^\a$.
 Similarly in such Fermi coordinates,  one may write the connection 1-forms in
this coframe  \be{\omega^\aa{}_\bb = \AA\aa{{\bb i}} d\sigma^i +
\CC\aa\bb\label{E3}} where $\AA\aa{{\bb i}}$ are 0 forms dependent
on $\sigma,^i\,\l,\,p^\a $ and $\CC\aa{{\bb i}}$ are 1-forms
independent of $d\sigma^i ,\, d\l$ but with components dependent
on $\sigma^i,\,\l,\,p^\a$. All the  functions appearing in these
expressions can be found as before by solving the structure
equations that define the torsion and curvature of $\nabla$.

The source currents $\tau_\aa$, $S^{\aa\bb}$ and $j$   are now
localised on the tubular region $\UU$ about the rod history $\s$.
The parameterisation of the stress-energy 3-form currents
$\t^\aa$, $S^{\aa\bb}$ on $\s$ defines the rod variables used to
pass from a field to a {\lq\lq{} {Cosserat rod} \rq\rq}
description \footnote{Strings arise as a special case of
shear-free rods}.
 In the absence of dissipation one can generate relativistic
 Cosserat models interacting with gravitational and electromagnetic fields from:
\be{\t^\aa=P^\aa{}_i\se i + \chi F^{\aa\bb}\,\SS_{\bb\bbb}\se
\bbb\label{taucurrent}} \be{S_\aa{}^\bb=\SS{}_\aa{}^\bb{}_j \se
j\label{spincurrent}} \be{j=\rho\se 0+ J\se1.\label{emcurrent}} in
terms of the scalar components $P^\aa{}_i$, $\Sigma^{\aa\bb}{}_j$,
$F^{AB}$,  $\rho,\,J$ on $\UU$ and constant $\chi$. For thermally
active media one must supplement these equations with
phenomenological consitutive relations for the stress-energy,
spin, entropy and heat forms subject to the constraints
 imposed by entropy production.

\section{Conclusions}

With the aid of the structure equations for a metric-compatible
connection on the bundle of analytic orthonormal frames over a
tubular domain of  spacetime, a  method for constructing a local
section by Taylor series has been described. Explicit computations
have been presented in terms of Fermi coordinates for this domain
associated with a (generalised) Fermi-Walker frame attached to a
timelike curve. This tubular section is used to effect a
truncation scheme based on balance laws associated with identities
derived from the invariances of an action integral. The curve is
defined as the history of a massive particle with spin. A series
of auxiliary {\it constitutive} relations is used to construct
models for the stress-energy and electromagnetic currents that
enter into these balance laws. Taylor series representations are
expressed in terms of  expansions in  the arc distance along
spacelike autoparallels associated with the Fermi-Walker frame. By
suitably truncating these series one may construct models of
spinning matter. In the absence of self-forces the determination
of a consistent dynamical scheme for the acceleration of the
particle worldline can be effected by ensuring that the auxiliary
conditions render each truncated set an involutive system of
ordinary differential equations. This procedure has been
demonstrated by analysing the dynamical equations in Minkowski
spacetime where it becomes possible to identify the expected
interactions of a charged spinning particle with a particular
magnetic moment in an external electromagnetic field.

A number of interesting features emerges from this investigation.
The inclusion of charged matter interactions with gravity yields
terms that may be interpreted as tidal forces that couple
components of the torsion tensor to the electromagnetic field via
the particle spin. Such interactions are novel and may offer new
experimental probes for the detection of spacetime torsion.
Another feature is that in the lowest approximation considered
here, electrically neutral massive {\it spinless} particles are
predicted to follow {\it autoparallels} of the Levi-Civita
connection. By contrast the precession rate and motion of spinning
gyroscopes in gravitational fields with torsion differ from those
in a torsion-free environment \CITE{rwt_clark}. The differences
can be estimated from the results given in this paper.

 The approach adopted here offers a number of natural
generalisations. Fermi charts adapted to spacetime submanifolds
with co-dimension less than three permit the methods to be applied
to derive the relativistic motion of Cosserat media \cite{Epstein:
2001} in general spacetimes. The associated orthonormal coframe
and source currents for  a relativistic rod have been mentioned in
$IX$.

The above analysis has been conservative. Prompted by puzzles in
cosmology many modern theories of gravitation invoke   scalar
fields that couple directly to spacetime geometry and Yang-Mills
fields. Some of these approaches suggest that at some scale the
dilation group may arise as an additional local symmetry. If this
group is included it is possible to generalise the balance laws to
incorporate fields with Weyl charge in non-Riemannian geometries.
It is  then possible that not all electrically neutral {\it
spinless} particles  have the same type of worldline.  In
geometries with torsion, autoparallel worldlines are in general
distinct from geodesics of the (torsion-free) Levi-Civita
connection \CITE{geroch}. Thus the shift in perihelia of celestial
objects in highly eccentric orbits may also be used as probes in
generalised theories of gravitation \CITE{dereli_tucker_tor}.

A discussion of the significance of radiative forces on the motion
of charged particles in geometries with torsion and an extension
of the methods to retarded null coordinates associated with an
arbitrary timelike curve will be presented elsewhere.

\section{Acknowledgements}

The author is most grateful to Tekin Dereli, David Burton, Miguel
S$\acute a$nchez, Charles Wang, and  Ron Evans for useful
conversations and to
 BAe Systems for supporting this research. Particular thanks are also due to
 Adam
 Noble for a careful reading of the manuscript.





\end{document}